\documentclass[manuscript]{acmart}

\usepackage{microtype}
\usepackage{booktabs}
\usepackage{graphicx}
\usepackage{subcaption}
\usepackage{array}
\usepackage{tabularx}
\usepackage{float}
\usepackage{multirow}
\setcopyright{acmlicensed}
\acmConference[IUI '27]{ACM International Conference on Intelligent User Interfaces}{February 8--11, 2027}{Helsinki, Finland}
\acmYear{2027}
\acmDOI{}

\begin{document}

\title{Memdora: Designing Cognitively-Grounded Flashcard Interactions for AI-Powered Spaced Repetition}

\author{Ruiyang Zhang}
\affiliation{\institution{Ryonix Labs Inc.}\city{Montreal}\state{Quebec}\country{Canada}}
\email{zhangruiyang36@gmail.com}

\begin{abstract}
Spaced repetition systems (SRS) have demonstrated robust effects on long-term retention, yet existing tools reduce the flashcard interaction to a single binary gesture: flip and self-rate. This impoverished interaction model fails to leverage decades of cognitive science evidence on retrieval practice, and requires learners to context-switch out of their reading flow to create cards manually. We present Memdora, a cross-platform AI spaced repetition system that addresses these limitations through four contributions: (1) a taxonomy of 17 cognitively-grounded interaction types across three learning categories---Language (6~types), By Heart (1~type with 3~retrieval modes), and Exam (10~types)---each grounded in peer-reviewed cognitive science evidence, with per-type design rationale and citations documented in this paper; (2) a unified AI generation pipeline that collapses card creation to a single gesture at the point of reading across web, mobile, and three browser extensions (Chrome, Edge, Firefox); (3) a collaborative layer enabling users to publish decks with live synchronization: followers discover and follow decks via a public feed, and any edits the deck owner makes propagate instantly to all followers while each follower maintains independent FSRS-6 scheduling state; and (4) an effort-based behavioral reward system that incentivizes actual cognitive engagement rather than mere app presence. Memdora integrates FSRS-6, the current state-of-the-art spaced repetition algorithm, and is deployed publicly on iOS, Android, Web, and three browser extensions. We describe the design rationale for each interaction type, discuss how the system advances beyond prior AI flashcard systems including SmartFlash and KARL, and outline implications for educational technology design.
\end{abstract}




\maketitle

%%-------------------------------------------------------
\section{Introduction}

The forgetting curve, first described by Ebbinghaus in 1885~\cite{ebbinghaus1885}, remains one of the most replicated findings in cognitive psychology: without review, humans forget approximately 70\% of newly learned material within 24~hours. Spaced repetition systems (SRS) address this by scheduling reviews at optimally expanding intervals, leveraging the spacing effect to dramatically improve long-term retention~\cite{roediger2006}.

Despite decades of evidence supporting spaced repetition, mainstream SRS tools---most notably Anki, with over 10~million downloads---have changed little since the introduction of SuperMemo's SM-2 algorithm in 1987. Critically, these tools reduce the flashcard interaction to a single gesture: flip the card, reveal the answer, self-rate recall. This binary interaction model has two significant limitations.

First, it ignores decades of cognitive science evidence on retrieval practice. Research demonstrates that different retrieval formats---multiple choice, free recall, fill-in-the-blank, error detection, sequential reconstruction---activate different cognitive processes and produce differential retention benefits~\cite{marsh2007,morris1977}. A vocabulary learner who always flips a card to see a translation is engaging in recognition-based recall, among the weakest forms of retrieval~\cite{marsh2007}. More effortful retrieval formats consistently produce superior long-term retention through the testing effect~\cite{roediger2006}.

Second, existing tools impose significant creation friction. Anki users must manually type each card, switching context from their reading material to the application. This preparation burden deters sustained adoption: a recent study found that students reported spending more energy preparing cards than actually studying them~\cite{smartflash2026}.

\begin{figure}[htbp]
  \centering
  \includegraphics[width=0.95\textwidth]{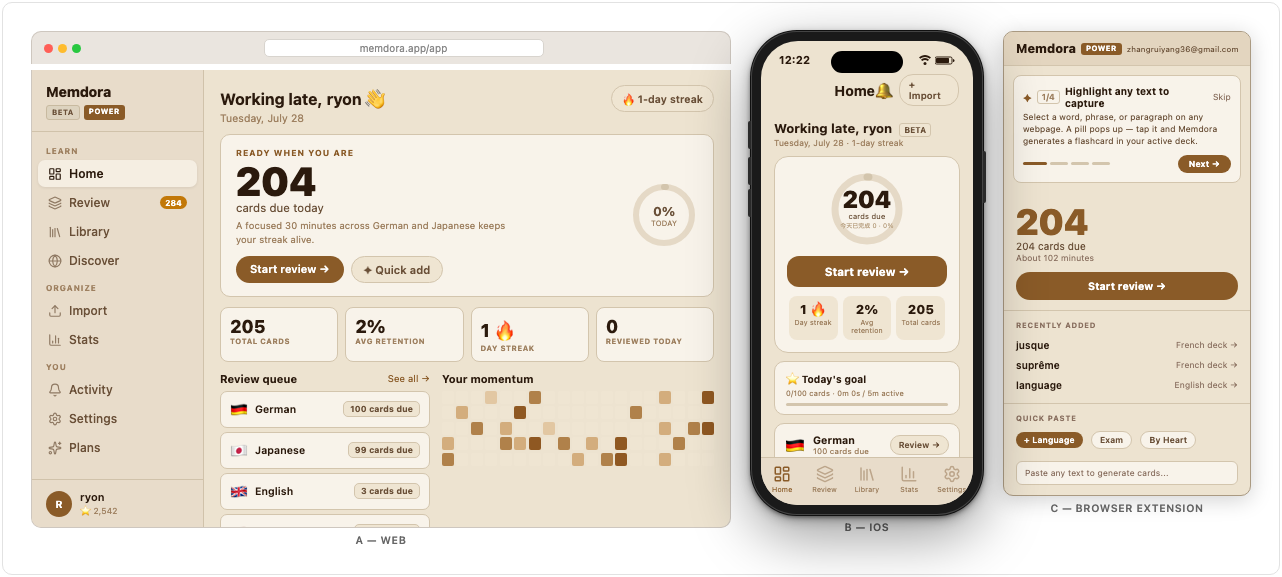}
  \caption{Memdora deployed across three surfaces sharing a single backend. (A)~Web dashboard showing the review queue, per-deck due counts, and daily momentum heatmap. (B)~iOS home screen with circular progress ring, multi-deck queue, and today's goal tracker. (C)~Browser extension popup with in-context capture onboarding, due-card count synced across surfaces, and quick-paste card generation.}
  \label{fig:overview}
\end{figure}

We present Memdora, an AI-powered spaced repetition system (Figure~\ref{fig:overview}) that addresses both limitations through four primary contributions:
\begin{enumerate}
  \item A taxonomy of 17 cognitively-grounded interaction types across three learning categories (Table~\ref{tab:taxonomy}), each grounded in peer-reviewed cognitive science---with per-type citations and design rationale documented in this paper.
  \item A unified AI generation pipeline that collapses card creation to a single gesture at the point of reading across web, mobile, and three browser extensions.
  \item A collaborative layer enabling users to publish decks with live synchronization: followers discover and follow decks via a public feed, and any edits the deck owner makes propagate instantly to all followers while each follower maintains independent FSRS-6 scheduling state.
  \item An effort-based behavioral reward system that incentivizes actual cognitive engagement (5~minutes of study time, 100~cards reviewed) rather than mere app presence.
\end{enumerate}

Memdora is deployed publicly on iOS, Android, Web, and browser extensions for Chrome, Edge, and Firefox. It integrates FSRS-6~\cite{fsrs2022}, the current state-of-the-art open-source spaced repetition algorithm, which outperforms Anki's SM-2 in recall prediction accuracy. Mobile apps support offline review with background synchronization.

%%-------------------------------------------------------
\section{Related Work}

\subsection{Spaced Repetition Systems}

Spaced repetition builds on the spacing effect---the finding that distributed practice over time produces better long-term retention than massed practice~\cite{roediger2006}. The SuperMemo SM-2 algorithm (1987) operationalized spaced repetition as a card scheduling system and became the basis for Anki. FSRS~\cite{fsrs2022}, currently in version~6, uses a differential equation model of memory stability and retrievability, achieving significantly lower mean absolute error in recall prediction than SM-2. KARL~\cite{karl2024} further advanced scheduling by incorporating card content through BERT embeddings, demonstrating that semantic relationships between cards can improve scheduling accuracy on a dataset of 123,143~study logs.

Despite these algorithmic advances, the interaction layer of SRS tools has remained largely unchanged. Neither Anki, Quizlet, nor existing research systems have proposed a systematic, cognitively-grounded taxonomy of flashcard interaction types.

\subsection{AI-Generated Flashcards}

SmartFlash~\cite{smartflash2026} studied AI flashcard generation using LLMs and found that learners valued automation for reducing preparation burden, but required editable, transparent AI outputs to maintain cognitive ownership. The study involved 6~students using a research prototype with a single basic card type. Bachiri et al.~\cite{bachiri2023} demonstrated comparable quality between AI-generated and human-created flashcards in MOOC contexts. Agnes and Srinivasan~\cite{agnes2024} found that AI-generated mnemonic keywords significantly improved vocabulary retention over standard flashcards.

Memdora extends this line of work by deploying AI generation across five platforms with a single-gesture capture interface, and generating cards across 17~distinct cognitively-grounded interaction types.

\subsection{Retrieval Practice and Interaction Format}

Retrieval practice consistently produces superior retention compared to passive re-reading~\cite{roediger2006}. Critically, retrieval format matters: multiple-choice produces recognition-based retrieval, while free recall requires generative retrieval---a more effortful process producing stronger memory traces~\cite{marsh2007}. Elaborative interrogation produces strong deep encoding~\cite{graesser2003}. Bjork's desirable difficulties framework~\cite{bjork1994} establishes that varied, effortful retrieval conditions consistently produce better long-term retention, motivating Memdora's interaction diversity and the By Heart category's self-regulated difficulty modes.

\subsection{Gamification and Behavioral Retention}

Duolingo's gamification research demonstrates that streaks and reward systems improve daily retention and engagement~\cite{duolingo2023}. However, presence-based rewards may produce superficial engagement rather than deep learning~\cite{hamari2014}. Memdora rewards effort rather than presence: users earn AI generation credits for completing 5~continuous minutes of study and for reviewing 100~cards per day.

%%-------------------------------------------------------
\section{System Overview}

Memdora is a cross-platform SRS available as a web application, iOS and Android apps, and browser extensions for Chrome, Edge, and Firefox (Figure~\ref{fig:overview}). The system comprises four layers: capture, generation, scheduling, and review.

\subsection{Capture Layer}

\begin{figure}[htbp]
  \centering
  \includegraphics[width=0.95\textwidth]{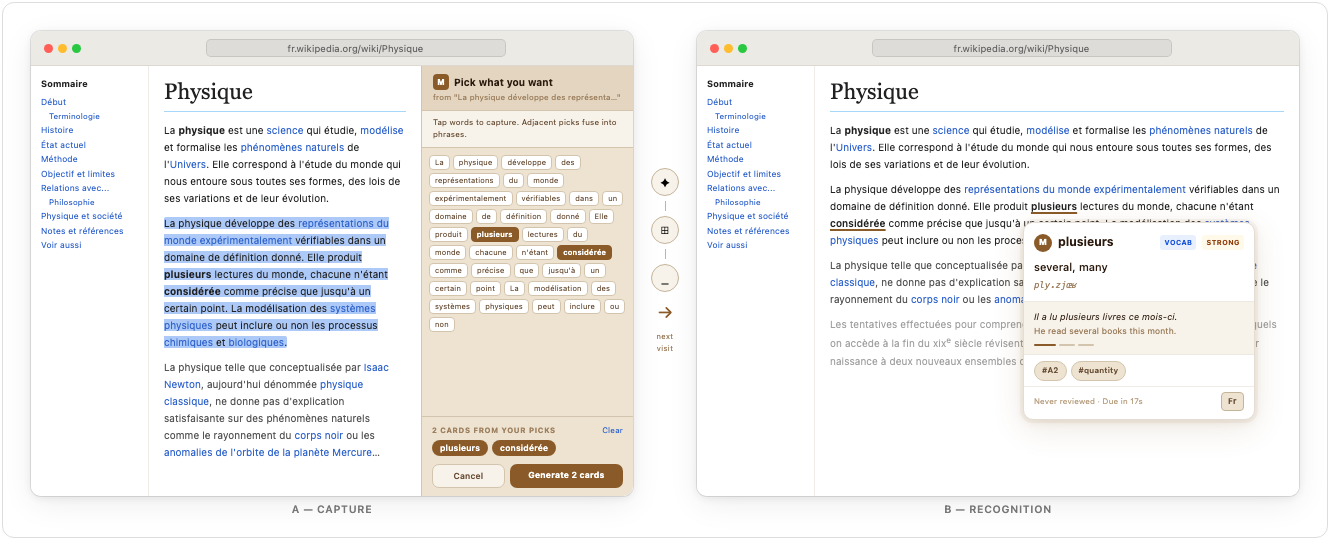}
  \caption{The browser extension operates bidirectionally on any webpage. (A)~Capture: selected text triggers a word chip picker; tapping chips queues them for AI card generation without leaving the page. (B)~Recognition: hovering over a word already in the user's deck surfaces the card inline---showing type badge, retention percentage, and example sentence---enabling incidental spaced retrieval during normal reading.}
  \label{fig:extension}
\end{figure}

Users add content through four modalities: (a)~pasting an article or YouTube URL, (b)~uploading a PDF, (c)~direct text entry, or (d)~using the browser extension to highlight text on any web page. The browser extensions (Figure~\ref{fig:extension}) provide single-gesture capture: the user highlights text and clicks the extension popup to immediately trigger AI card generation. This collapses the context-switching cost identified in SmartFlash~\cite{smartflash2026} to near zero.

The extension also supports \textit{in-context vocabulary recall}: when a user hovers over a word on any webpage that already exists in their Memdora deck, a tooltip popup displays the card---including its current FSRS-6 retention percentage, card type, and example sentence---without requiring the user to open the application. This bidirectional design transforms the browser into both a capture surface and a passive review surface, enabling incidental spaced retrieval during normal reading sessions.

\subsection{AI Generation Layer}

\begin{figure}[htbp]
  \centering
  \includegraphics[width=0.95\textwidth]{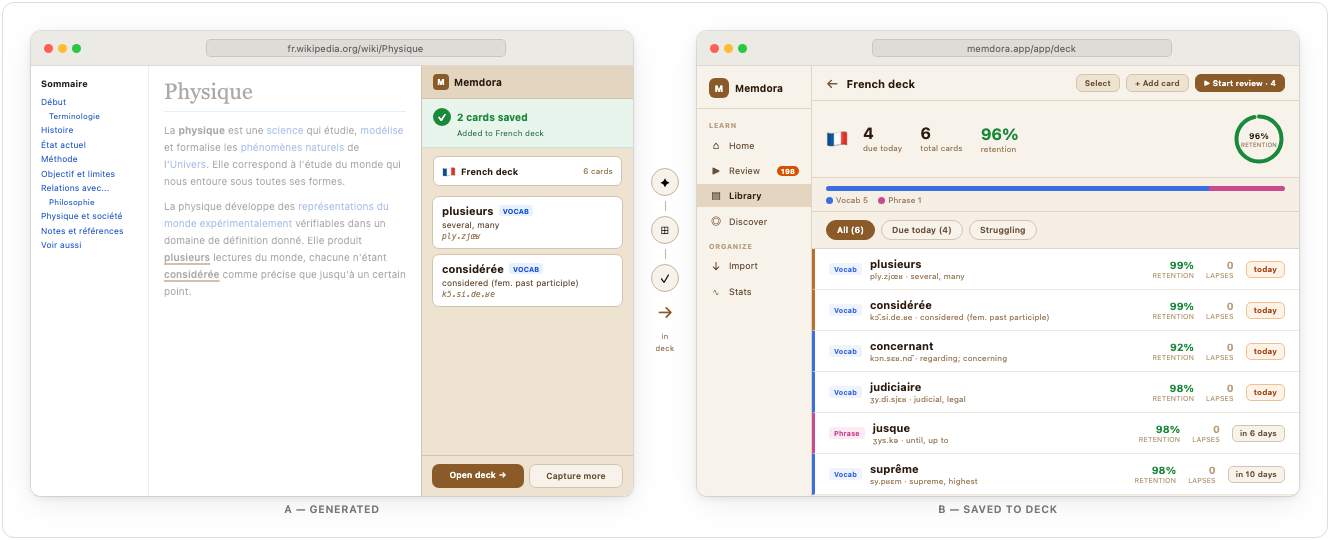}
  \caption{The AI generation pipeline: (A)~A confirmation popup appears immediately after card capture, displaying generated card content including phonetic transcriptions and translations; cards are automatically saved to the active deck. (B)~The deck browser shows all cards with per-card FSRS-6 retention percentages, lapse counts, and next review dates, giving learners full visibility into their memory state.}
  \label{fig:generation}
\end{figure}

The AI generation pipeline uses a large language model with structured output prompting. Content-type classification routes extracted knowledge items to appropriate interaction type(s): vocabulary items to Language cards, quotable passages to By Heart cards, factual content to Exam cards. For multiple-choice formats, the pipeline generates three plausible distractor options drawn semantically from the same deck. Generated cards are immediately saved to the active deck and shown in a confirmation popup (Figure~\ref{fig:generation}); they remain editable at any time in the deck browser, consistent with SmartFlash's finding that editability is essential for learner cognitive ownership~\cite{smartflash2026}.

\subsection{Scheduling Layer}

\begin{figure}[htbp]
  \centering
  \includegraphics[width=0.95\textwidth]{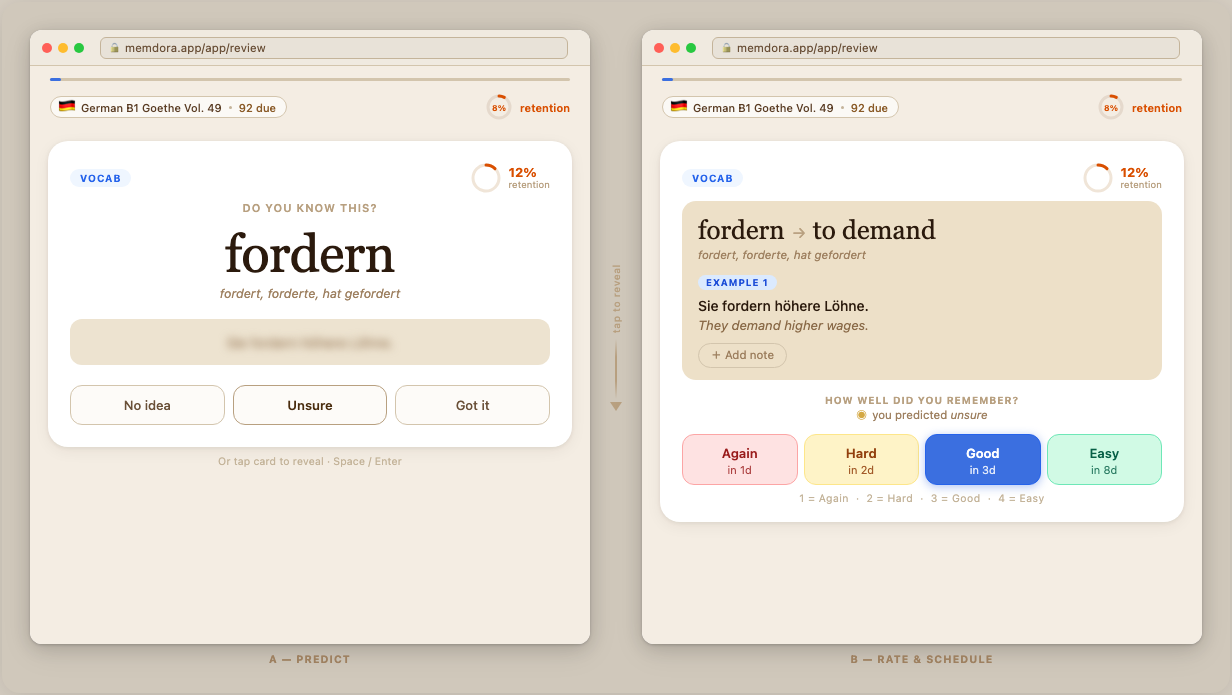}
  \caption{The review interaction for a Vocabulary card. (A)~Predict: the card displays the word, phonetic conjugation, and current FSRS-6 retention percentage (top right); the learner self-predicts recall before revealing the answer. (B)~Rate \& Schedule: after reveal, the learner rates difficulty; FSRS-6 computes the next review interval for each rating (Again~1d, Hard~2d, Good~3d, Easy~8d).}
  \label{fig:card}
\end{figure}

All cards are scheduled using FSRS-6~\cite{fsrs2022}, which maintains per-card estimates of memory stability ($S$) and retrievability ($R$). The system computes optimal review intervals to maintain target retention (default~90\%). FSRS-6 runs client-side on mobile for offline scheduling, with server-side synchronization across platforms. Every card displays its current retention percentage to the learner during review (Figure~\ref{fig:card}), building metacognitive awareness of memory state.

\subsection{Collaborative Features}

\begin{figure}[htbp]
  \centering
  \includegraphics[width=0.95\textwidth]{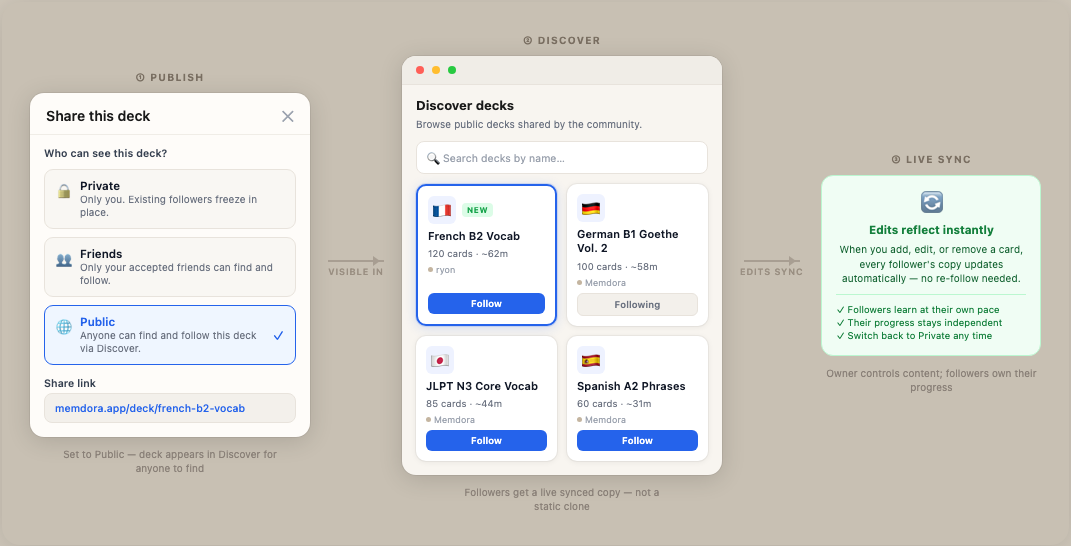}
  \caption{The collaborative deck layer. (Publish)~Deck owners set visibility to Private, Friends, or Public, generating a shareable link. (Discover)~Any user can browse community decks by language and follow with one tap. (Live Sync)~When the deck owner edits, adds, or removes cards, all followers' copies update instantly without requiring a re-follow; each follower's review progress remains independent.}
  \label{fig:classroom}
\end{figure}

The collaborative deck layer (Figure~\ref{fig:classroom}) allows users to publish decks publicly, to friends, or keep them private. Published decks appear in the Discover feed, where any user can browse and follow them. Crucially, when the deck owner adds, edits, or removes a card, all followers' copies update instantly---no re-follow required. Each follower maintains independent FSRS-6 scheduling state, learning at their own pace. This living deck model supports classroom use: a teacher publishes a shared vocabulary deck, students follow it, and the teacher can push new cards or corrections that propagate to all students automatically.

%%-------------------------------------------------------
\section{A Taxonomy of 17 Cognitively-Grounded Interaction Types}

The central design contribution of Memdora is its taxonomy of 17~interaction types across three learning categories (Table~\ref{tab:taxonomy}). Each type is grounded in peer-reviewed cognitive science research, as documented in Table~\ref{tab:taxonomy}.

\subsection{Design Philosophy}

The taxonomy is built on three principles. First, retrieval effort produces retention: more effortful retrieval formats produce stronger memory traces~\cite{roediger2006}. Second, format should match learning goal: vocabulary acquisition, verbatim memorization, and conceptual understanding require different retrieval formats~\cite{marsh2007}. Third, transparency builds metacognition: making algorithmic memory predictions visible promotes learner trust and self-regulation~\cite{smartflash2026}.

\subsection{Language Category (6~Types)}

\begin{figure}[htbp]
  \centering
  \includegraphics[width=0.95\textwidth]{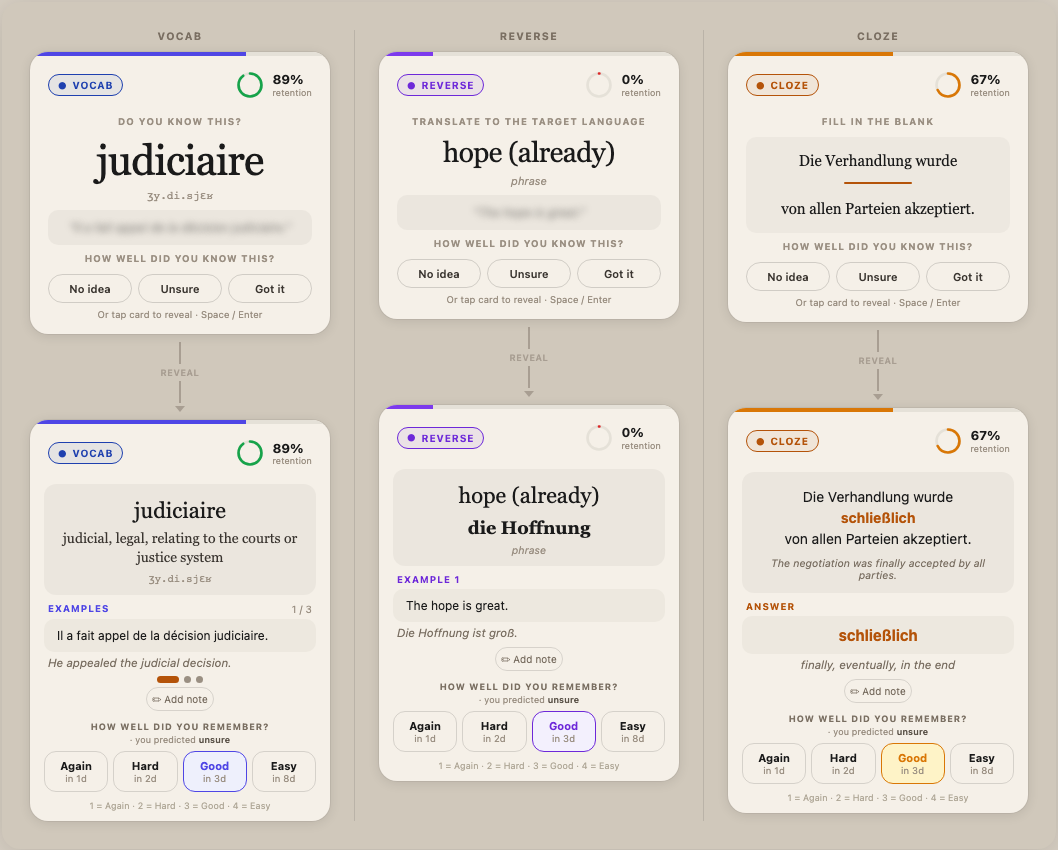}
  \caption{Three of the six Language interaction types (before and after reveal). Left: Vocabulary---self-rate recall of a French word before revealing translation, phonetics, and contextual examples. Center: Reverse---produce the target-language form from a source-language phrase prompt. Right: Cloze---supply the missing word in a German sentence before the answer is revealed.}
  \label{fig:language}
\end{figure}

The Language category (Figure~\ref{fig:language}) supports vocabulary and phrase acquisition in a target language.

\textbf{Vocabulary.} Target-language word with phonetic transcription and source sentence; learner self-rates recall before revealing translation. Grounded in retrieval practice~\cite{roediger2006}.

\textbf{Cloze.} Sentence with one key word blanked; learner supplies the missing word. Produces stronger retention than translation recall~\cite{bransford1972}.

\textbf{Phrase.} Multi-word expression in context requiring production of full meaning. Grounded in phrasal chunking research~\cite{sinclair1991}.

\textbf{Concept.} Learner explains a concept before revealing the target explanation. Grounded in elaborative interrogation~\cite{graesser2003}.

\textbf{Reverse.} Definition presented in native language; learner produces the target-language form. Bidirectional retrieval strengthens encoding~\cite{richland2005}.

\textbf{Sentence.} Context sentence with key phrase highlighted; learner judges usage correctness. Grounded in contextual learning~\cite{sinclair1991}.

\subsection{By Heart Category (1~Type, 3~Retrieval Modes)}

\begin{figure}[htbp]
  \centering
  \includegraphics[width=0.95\textwidth]{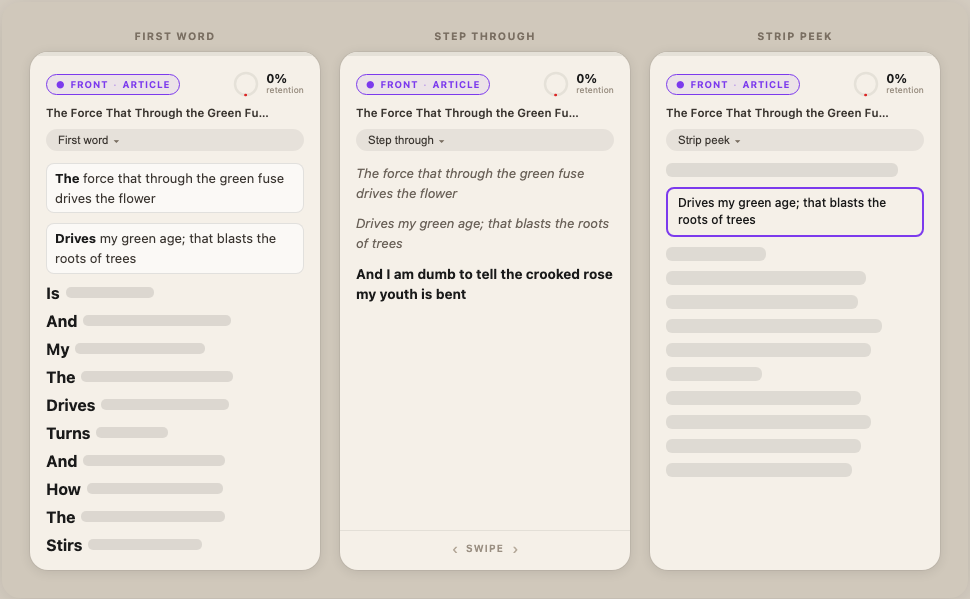}
  \caption{The three By Heart retrieval modes applied to the same poem (Dylan Thomas, \textit{The Force That Through the Green Fuse}). Left: First Word---only the opening word of each line is visible as a retrieval cue. Center: Step Through---past lines are dimmed; the current target line is bold; learner swipes to advance. Right: Strip Peek---all lines are hidden; learner taps any line to reveal it for 2~seconds. Difficulty increases left to right; learners self-select mode per session.}
  \label{fig:byheart}
\end{figure}

The By Heart category (Figure~\ref{fig:byheart}) supports verbatim memorization of extended passages. Unlike other categories, By Heart presents a single card type with three user-selectable retrieval modes of varying difficulty, grounded in Bjork's desirable difficulties framework~\cite{bjork1994} and self-regulated learning research~\cite{pintrich2000}.

\textbf{Strip Peek.} All lines hidden; learner taps any line to reveal it for 2~seconds. Maximum active recall---highest-effort mode.

\textbf{Step Through.} Past lines dimmed; current target line bold. Provides partial cueing from preceding context.

\textbf{First Word.} Each line represented by its opening word only. Lightest cue, most support.

The three modes constitute three points on the cue-target continuum identified in retrieval practice research. Allowing learners to self-select difficulty on a per-session basis operationalizes the desirable difficulties framework within a practical UI---to our knowledge, the first SRS system to do so for extended passage memorization.

\subsection{Exam Category (10~Types)}

\begin{figure}[htbp]
  \centering
  \includegraphics[width=0.95\textwidth]{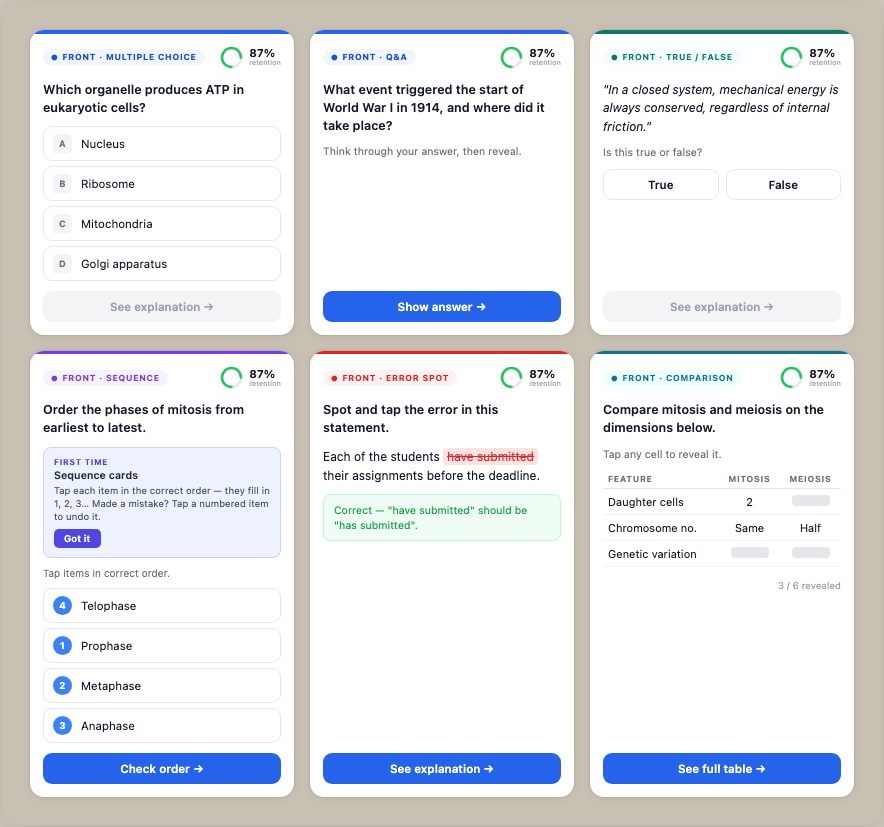}
  \caption{Six of the ten Exam interaction types. Top row: Multiple Choice (select from 4 options with AI-generated distractors), Q\&A (free-recall before reveal), True/False (binary judgment of a factual statement). Bottom row: Sequence (order steps or events), Error Spot (identify and correct a deliberate error in a statement), Comparison (articulate relationships via a reveal table).}
  \label{fig:exam}
\end{figure}

The Exam category (Figure~\ref{fig:exam}) covers the primary retrieval formats in educational assessment research (Table~\ref{tab:taxonomy}).

\begin{table*}[t]
\centering
\caption{Memdora's taxonomy of 17 cognitively-grounded interaction types. Each type is paired with its primary cognitive science grounding citation.}
\label{tab:taxonomy}
\begin{tabularx}{\textwidth}{llXl}
\toprule
\textbf{Category} & \textbf{Type} & \textbf{Interaction Description} & \textbf{Primary Citation} \\
\midrule
\multirow{6}{*}{Language}
 & Vocabulary    & Self-rate recall before revealing translation & Roediger \& Karpicke~\cite{roediger2006} \\
 & Cloze         & Supply missing word in target-language sentence & Bransford \& Johnson~\cite{bransford1972} \\
 & Phrase        & Produce meaning of multi-word expression in context & Sinclair~\cite{sinclair1991} \\
 & Concept       & Explain concept before revealing target explanation & Graesser \& Olde~\cite{graesser2003} \\
 & Reverse       & Produce target-language form from native-language prompt & Richland et al.~\cite{richland2005} \\
 & Sentence      & Judge usage correctness of highlighted phrase & Sinclair~\cite{sinclair1991} \\
\midrule
\multirow{3}{*}{By Heart}
 & Strip Peek    & Tap to reveal each line for 2 seconds (highest effort) & Bjork~\cite{bjork1994} \\
 & Step Through  & Current line bold; past lines dimmed (medium effort) & Bjork~\cite{bjork1994} \\
 & First Word    & Opening word only as retrieval cue (lowest effort) & Pintrich~\cite{pintrich2000} \\
\midrule
\multirow{10}{*}{Exam}
 & Multiple Choice & Select from 4 options (3 AI-generated distractors) & Marsh et al.~\cite{marsh2007} \\
 & Q\&A            & Free-recall response before revealing answer & Roediger \& Karpicke~\cite{roediger2006} \\
 & True/False      & Binary judgment of factual statement & Marsh et al.~\cite{marsh2007} \\
 & Matching        & Pair related terms or concepts & Richland et al.~\cite{richland2005} \\
 & Sequence        & Place steps or events in correct order & Sinclair~\cite{sinclair1991} \\
 & Formula         & Complete mathematical or chemical formula & Sweller~\cite{sweller1988} \\
 & Scenario        & Apply knowledge to realistic context & Morris et al.~\cite{morris1977} \\
 & Error Spot      & Identify and correct deliberate error in statement & Marsh et al.~\cite{marsh2007} \\
 & Comparison      & Articulate difference or relationship between two concepts & Graesser \& Olde~\cite{graesser2003} \\
 & Fill the Blank  & Supply key term removed from sentence & Bransford \& Johnson~\cite{bransford1972} \\
\bottomrule
\end{tabularx}
\end{table*}

%%-------------------------------------------------------
\section{Behavioral Design}

\subsection{Effort-Based Reward System}

Memdora rewards cognitive effort through two primary mechanisms. The \textbf{Study Time Reward} grants bonus AI generation credits for completing 5~continuous minutes of active card review. The \textbf{Volume Reward} grants additional credits for reviewing 100~cards in a day. A \textbf{Streak System} tracks consecutive days of actual card review (not mere app opening), providing continuity motivation through loss aversion~\cite{duolingo2023}. This effort-based design contrasts with presence-based rewards (rewarding any app interaction) shown to produce superficial engagement~\cite{hamari2014}.

\subsection{Transparency Features}

Every card displays the current retention percentage computed by FSRS-6 in the top-right corner (Figure~\ref{fig:card}), making the scheduling model's memory predictions visible during review. The deck browser additionally shows per-card retention estimates alongside next review dates, enabling learners to identify their weakest cards at a glance. These features build metacognitive awareness of one's own memory state, responding directly to SmartFlash's finding that transparency in AI educational systems is essential for learner acceptance~\cite{smartflash2026}.

%%-------------------------------------------------------
\section{Implementation}

Memdora is implemented as a full-stack TypeScript application. The web frontend uses React with server-side rendering. Native mobile apps use React Native (Expo) for iOS and Android. Browser extensions use the WebExtensions API, providing compatibility across Chrome, Edge, and Firefox from a shared codebase. The AI generation pipeline uses a large language model with typed JSON output schemas corresponding to each of the 17~interaction types.

FSRS-6 runs client-side against a local SQLite database on mobile, enabling fully offline scheduling, with server-side synchronization ensuring cross-platform consistency. The offline-first mobile architecture is critical for language learners who review during commutes without reliable connectivity.

The deck publishing infrastructure uses a publish-subscribe model. Followers receive live updates when the deck owner modifies content, with each follower maintaining independent scheduling state server-side.

%%-------------------------------------------------------
\section{Discussion}

\subsection{Design Implications}

The 17-type taxonomy raises an important design question: does interaction diversity improve or impair learning? Bjork's desirable difficulties research~\cite{bjork1994} suggests varied retrieval formats produce better long-term retention, as variation prevents format-specific learning strategies. However, excessive novelty may increase extraneous cognitive load~\cite{sweller1988}. Memdora addresses this by selecting interaction types based on content type rather than random variation, preserving format-meaning coherence while delivering retrieval format diversity.

The By Heart 3-mode design operationalizes a key insight from self-regulated learning research: learners benefit from agency over difficulty selection~\cite{pintrich2000}. Displaying all three options simultaneously makes the difficulty spectrum explicit, potentially improving learner metacognition about their own recall ability.

\subsection{Limitations}

The primary limitation is the absence of a controlled user study evaluating the interaction taxonomy's effectiveness. The design rationale for each interaction type is grounded in existing research, but the efficacy of these types within Memdora --- and in combination with FSRS-6 scheduling --- has not been empirically evaluated. We are designing a longitudinal study comparing retention outcomes across card types.

The AI generation pipeline may produce cards of variable quality in highly specialized domains. The collaborative deck features have not yet been evaluated in a formal educational setting; deployment may reveal unmet needs around teacher workflows and institutional authentication.

\subsection{Future Work}

We are designing Memdora Arena, a daily competitive challenge in which all users within a subject category receive the same AI-curated deck and compete on accuracy and speed against a global leaderboard. The Arena is designed to leverage social comparison for motivation while generating shareable result cards to drive organic user acquisition. Future work will evaluate whether competitive social features enhance or impede deep learning outcomes.

We also plan a formal classroom evaluation recruiting teachers across secondary and post-secondary institutions. The effort-based reward system will be evaluated in a separate study comparing learning depth metrics (retention at 30~days) between effort-rewarded and presence-rewarded conditions.

%%-------------------------------------------------------
\section{Conclusion}

We presented Memdora, a cross-platform AI spaced repetition system with four primary design contributions: a taxonomy of 17~cognitively-grounded interaction types spanning Language, By Heart, and Exam learning categories; a unified AI generation pipeline enabling single-gesture card creation across five platforms; a collaborative deck layer with live synchronization; and an effort-based behavioral reward system. The system integrates FSRS-6 for state-of-the-art scheduling and surfaces per-card retention percentages during review and in the deck browser.

The flashcard interaction has been systematically under-designed. The dominant flip-and-rate paradigm, unchanged since SuperMemo's SM-2 in 1987, ignores decades of evidence on retrieval practice format effects. By grounding 17~interaction types in peer-reviewed research and making FSRS-6 retention predictions visible during review, Memdora proposes that educational technology can be simultaneously evidence-based and transparent. We invite the IUI community to engage with the open questions this raises: how much interaction diversity is optimal, how learners self-regulate difficulty, and whether transparency about cognitive science mechanisms improves educational technology adoption and learning outcomes.

%%-------------------------------------------------------
\section*{GenAI Usage Disclosure}

Large language model tools were used in the preparation of this manuscript for the following purposes: (1)~grammar and prose editing of author-written text; (2)~generation of multiple-choice distractor options in the Exam card examples shown in Figure~7. All research design, system architecture, interaction taxonomy, and analysis were conducted by the authors. The core claims, contributions, and empirical content of this paper were not generated by AI tools.

%%-------------------------------------------------------
\bibliographystyle{ACM-Reference-Format}

\end{document}